# Comparing Correlation Lengths of Red and Blue Galaxies: A New Standard Length for Testing Cosmic Isotropy

Michael J. Longo[a]

Department of Physics, University of Michigan, Ann Arbor, MI 48109-1040

I introduce a simple empirical measure of average galaxy cluster sizes based on a comparison of the correlation lengths of red galaxies with blue that can provide a more accurate and bias-free measure of the average size and number density of galaxy clusters. Using 269,000 galaxies from the SDSS DR6 survey, I show that this 3D correlation length, averaged over many clusters, remains very nearly constant at $L_0$= 4.797±0.024 Mpc/$h$ from small redshifts out to redshifts of 0.5. This serves as a new measure of cosmic length scales and provides a means of testing the standard cosmological model that is almost free of selection biases. The unprecedented accuracy of the technique allows the possibility of sensitive searches for large-scale inhomogeneities. Applying the same technique to the Millennium Simulation galaxies I find very good agreement between it and the SDSS galaxies.

## 1. INTRODUCTION

"Old" red (mostly elliptical) galaxies are known to cluster more strongly than younger blue (spiral and irregular) galaxies (Zehavi et al. 2002, Skibba et al. 2008). This clustering reflects the density fluctuations in the early universe. At a particular redshift the length scales associated with this clustering are presumably constant everywhere and evolve slowly with time or redshift. I propose a new statistic based on the difference between the average correlation length of red and blue galaxies. As discussed below, the average 3D correlation length can be measured to a precision ≈ 2% out to $z \sim 0.5$, providing an order of magnitude improvement in accuracy over previous techniques. It provides another way to test the standard model and to probe for inhomogeneities.

Cluster sizes and counts are an important tool for understanding dark energy and other questions in cosmology (Smith et al. 2008; Takada 2006). Techniques using standard correlation functions are subject to serious biases because the density of observable galaxies decreases rapidly with increasing redshift (distance). Techniques that require explicit cluster finding are also subject to serious biases (Lima & Hu 2007). The technique described here does not require cluster finding, it automatically corrects for the decreasing density, and it appears to be almost bias free. The only assumption is that when averaged over sufficiently large volumes of space the clustering length of galaxies is constant or evolves slowly with $z$ (i.e., epoch) for $z < 1$. The very large data set allows a detailed study of systematic effects without the need to resort to simulations to correct for biases.

---

[a] mlongo@umich.edu

The SDSS DR6 database (Adelman-McCarthy et al.) contains ~680,000 galaxies with spectra for $z < 1.0$. The sky coverage of the spectroscopic data is almost complete for right ascensions ($RA$) between 120° and 250° and declinations ($\delta$) between -10° and 65°. For the hemisphere toward $RA=0°$, only narrow bands in $\delta$ near -10°, 0°, and +10° are covered. The following discussion will be limited to the hemisphere toward $RA=180°$ except as noted.

## 2. THE SAMPLE

All objects with spectra that were classified as "galaxies" in the SDSS DR6 database were used in this analysis. The definition of "red" and "blue" galaxies used here is based on the difference in luminosities in the ultraviolet ($U$) and far infrared ($Z$) optical bands (York et al. 2000). Figure 1 shows examples of the ($U$-$Z$) distribution for 3 of the redshift ranges. For small $z$ (Fig. 1a), the distribution shows two fairly well separated peaks corresponding to the bluer spiral and irregular galaxies on the left and the redder ellipticals on the right. For $z$~0.09 (Fig. 1b), the distribution has shifted somewhat toward the right due to overall reddening of the spectra, and the spiral peak is much diminished compared to the ellipticals. For larger $z$ (Fig. 1c), the distribution is dominated by a large peak near ($U$-$Z$)=0.5. This is due to the misidentification of quasars as "galaxies" in the sample. For large $z$, it is difficult to distinguish one from the other. The contamination of the galaxy sample with quasars was negligible for $z<0.3$. For $z>0.3$ the quasar positions were found to be correlated like blue galaxies, and they were therefore considered as part of the blue galaxy sample in determining correlation lengths.

To determine the correlation lengths of the elliptical (red) galaxies we can compare their correlation functions to those of the blue galaxies that do not exhibit strong clustering. Here the definition of red and blue is somewhat arbitrary. It is chosen empirically to maximize the correlations at small separations. For $z< 0.15$ the median $U$-$Z$ was used as the division between red and blue. Galaxies below the median were considered to be "blue" and those above "red". For higher redshifts for which the distinction between spirals and ellipticals is not apparent from the $U$-$Z$ plot, the division point was chosen to maximize the correlation amplitude at small separations. The divisions for the 3 examples in Fig. 1 are indicated.

The correlation lengths derived in what follows were found to be almost independent of the choice of the red-blue division. Errors due to the choice are discussed in the Uncertainties section. In the following I simply refer to the samples as "red" and "blue".

## 3. CORRELATION FUNCTIONS

Three-dimensional 2-point correlation functions for the galaxies in the red and blue samples were used to determine the cluster correlation lengths. For every pair of galaxies the dimensionless comoving coordinates of each galaxy were calculated. Assuming a Friedmann universe with deceleration parameter $q_0=0$, the comoving coordinates $s_1$ and $s_2$ are given by (Osmer 1981)

$$s_1 = z_1(1+\tfrac{1}{2}z_1)/(1+z_1), \qquad s_2 = z_2(1+\tfrac{1}{2}z_2)/(1+z_2) \qquad (1)$$

The comoving distance between the two galaxies is then the magnitude of the vector

$$\vec{r}\,' = (s_2\sin\theta, 0, s_2\cos\theta) + (0,0,-s_1)\left\{(1+s_2^2)^{\frac{1}{2}} + \frac{s_2}{s_1}\cos\theta - \frac{s_2}{s_1}\cos\theta(1+s_1^2)^{\frac{1}{2}}\right\} \qquad (2)$$

in units of $c/H_0$ where $H_0$ is Hubble's constant and $\theta$ is the angular separation of the two galaxies. In the following the separations $r$ will be calculated from Eq. 2. As discussed below, the correlation lengths are very insensitive to the choice of $q_0$.

I define the correlation functions used here as

$$\xi_{RR}(r) \equiv N_{RR}/\bar{N} - 1; \qquad \xi_{BB}(r) \equiv N_{BB}/\bar{N} - 1; \qquad \xi_{RB}(r) \equiv N_{RB}/\bar{N} - 1 \qquad (3)$$

where $N_{RR}$ is the number of red-red galaxy pairs in a particular separation bin at $r$; $N_{BB}$ is the number of blue-blue pairs; $N_{RB}$ is the number of red-blue pairs; and $\bar{N}$ is the mean of $N_{RR}$ and $N_{BB}$. Each of the $N$'s was normalized to the total number of pairs summed over all separation bins as defined below. The uncertainties in the $\xi$ are calculated as

$$\Delta\xi = \sqrt{N}/\bar{N} \cong 1/\sqrt{N} \qquad (4)$$

with $N$ being the unnormalized number of pairs defined above with the appropriate subscripts.

The correlations were studied in redshift slices between 0.01 and 0.75. For each $z$ range, the numbers of red and blue galaxies were first counted. Each of the red, blue, and (red+blue) samples was pared to $N_{small}$ galaxies, where $N_{small}$ is the smaller of red or blue, by choosing galaxies at random from among them. Then the separations, $r$, for all combinations of RR, BB, and RB galaxies were binned in sixty 0.00025 wide bins between $r=0$ and $r=0.015$. Occasionally in the SDSS data nearby galaxies appear more than once with different IDs when different points in the same galaxy are chosen as centers. Therefore pairs with separations $<1\times10^{-5}$ were excluded in order to remove possible duplicates. (This is several times the radius of a typical large galaxy.)

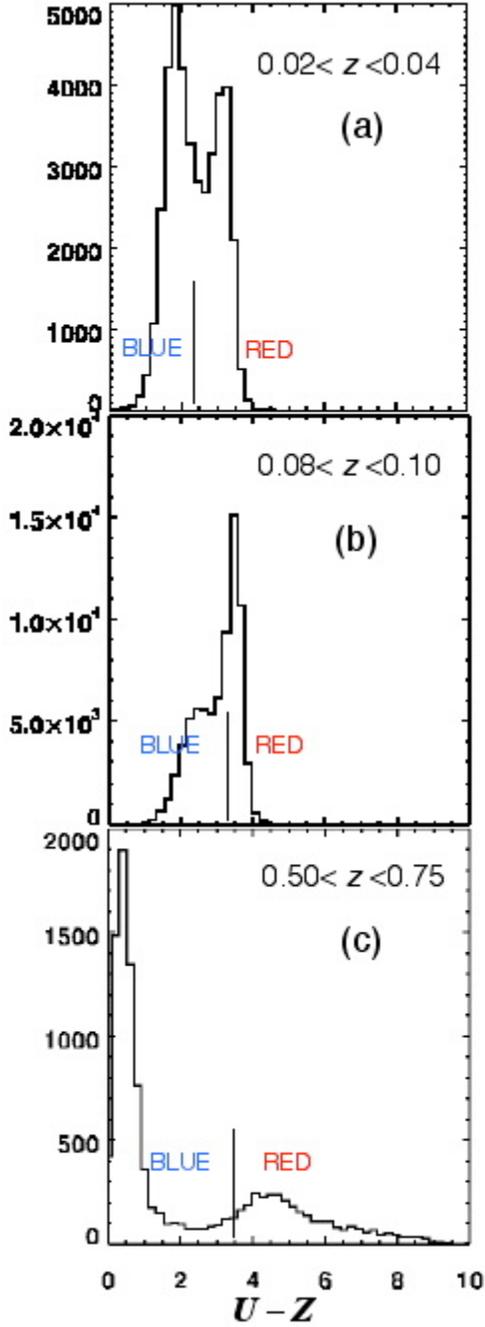
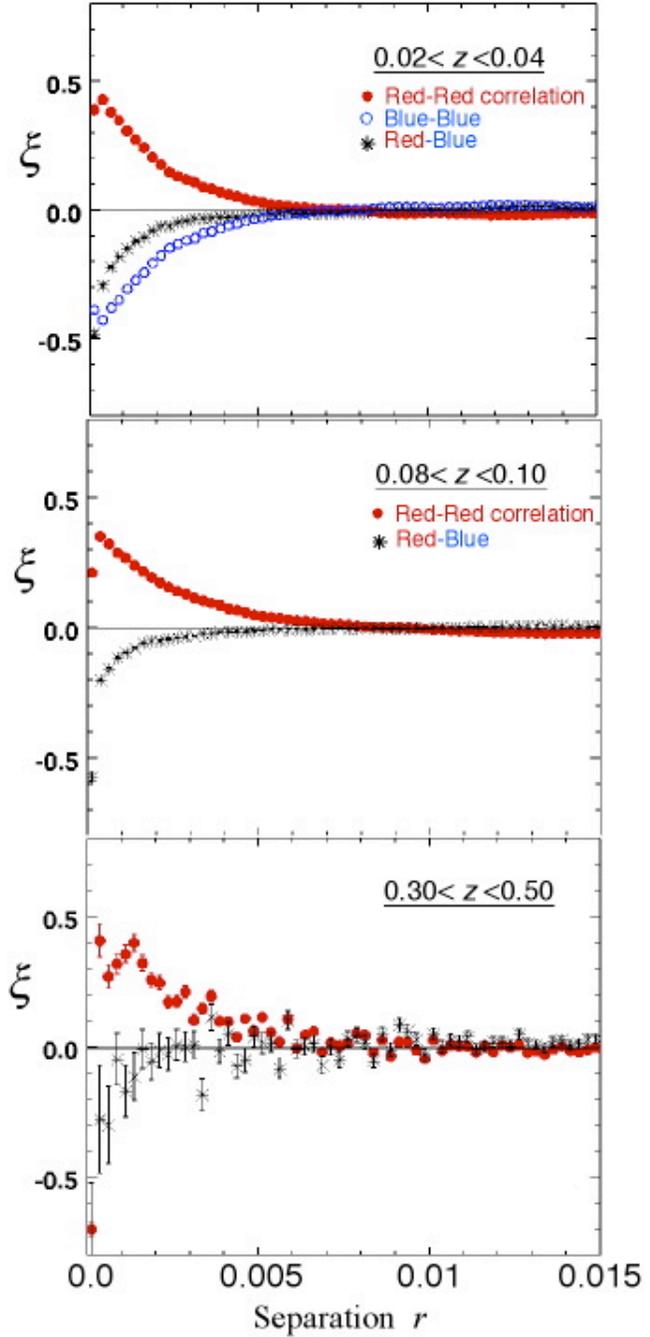

FIG. 1 *U-Z* distributions for 3 redshift ranges. The vertical lines show the division between "red" and "blue" galaxies.

FIG. 2 Correlation *vs.* separation for 3 redshift ranges. The red galaxies show a strong correlation at small separations and a strong anticorrelation with the blue. The error bars are usually smaller than the points.

The correlation functions were calculated as in Eq. (3) using the pair counts for each separation bin.

Typical examples of the correlations *vs.* separation are shown in Figure 2. The filled red circles are the $\xi_{RR}$, the blue open circles in Fig. 2a are the $\xi_{BB}$, and the asterisks are the $\xi_{RB}$. Error bars are usually smaller than the points. As defined in Eq. (3), $\xi_{RR}$ and $\xi_{BB}$ are mirror images of each other, so the $\xi_{BB}$ are only shown for the top plot. The red galaxies show a strong positive correlation with each other for separations <0.003 with an overwhelming statistical significance. The red and blue galaxies are very strongly anticorrelated with each other. For larger separations the $\xi$ are governed by the relative spatial densities of the red and blue galaxies in the chosen volume. For separations > 0.008 the correlations are roughly consistent with 0 over the whole redshift range studied. This shows that the technique of choosing equal numbers of red and blue galaxies in the volume works very well in removing systematic effects due to the spatial density of observed galaxies decreasing with distance.

The first separation bin is typically anomalous and was not used in the calculation of the correlation lengths. This is primarily due to the difficulty in resolving distinct galaxies that are close to each other on the sky and the masking of distant galaxies due to overlap with nearer ones. It has less effect on the red-blue correlation since it is easier to distinguish galaxies that are close to each other if they have different spectra. At larger $z$ there is some bias due to gravitational lensing but it affects the red and blue galaxies equally. For larger $z$ there is also some smearing of the $\xi$ at small separations due to the uncertainties in the SDSS redshift determinations. This was minimized by removing the few galaxies for which the uncertainty in $z$ is >0.02 $z$; it is only important for the larger $z$ slices. The contribution to the errors in the correlation lengths due to this smearing is discussed below.

There is no compelling model for the variation of $\xi$ with $r$. The definition of the correlation length $L_0$ is therefore somewhat arbitrary. As an *ad hoc* model-independent definition we use the mean $r$ over redshift bins 2 through 15 weighted by $\xi$ for that separation bin, $L_0 \equiv \sum_{2}^{15} \xi r / \sum_{2}^{15} \xi$.

This corresponds to separation bin centers in the range 0.000375 to 0.003875. The statistical uncertainties in the $L_0$ were calculated from the statistical errors in the 14 bins as given by Eq. (4). The statistical errors were usually much smaller than the total error (Table I). The length $L_0$ is a measure of the average radius of clusters in the redshift slice, as determined by the dark matter potential well the galaxies reside in.

## 4. UNCERTAINTIES AND SYSTEMATICS

The correlation lengths were found to be very robust relative to changes in the selection criteria. Many checks of the code were made. In particular, when the redshifts were scrambled at random among the galaxies the correlations disappeared entirely.

Due to the large number of galaxies in the sample, it was possible to study possible systematic effects and estimate the resulting uncertainties using the data itself. Generally the uncertainty estimates themselves were limited by the statistical error in the determination of $L_0$. These uncertainties are discussed in the following. They were combined in quadrature with the statistical error to get the total error given in Table I.

*Red-Blue Separation*.–The *U-Z* values for the red-blue separation are given in Table I. The uncertainties were estimated by varying the cut by at least 10% from the value there. They were comparable to the statistical errors with no apparent systematic effect with increasing *z*.

*Smearing due to zerr*.– The smearing in separations due to the uncertainty in the measured *z* tends to decrease $\xi$ at small *r*. This effect was studied by smearing the separations by an additional ±50% of *zerr*. This procedure could not be used for the 3 highest *z* bins because the statistical errors in the $\xi$ were too large; instead the additional smearing was varied until $\xi$ in the second bin decreased to half its maximum value. This error was always less than the statistical.

*Edge effects*.– Galaxies at the edge of the defined volumes, both the red and blue, will have neighbors on one side only. Possible edge effects were eliminated by looking at correlations between galaxies in a "box" and those in a larger box that was at least 0.005 larger on all sides. As a check, when the box sizes were made the same, there was no change in the $L_0$ within statistical errors. This was not possible with the 0.01<*z*<0.02 slice which is effectively all "edge". The uncertainty in $L_0$ for this slice was estimated as ±0.12 by varying the boundaries of the two boxes; this was the dominant error for that slice.

*Variation of $L_0$ with absolute magnitude*.– A galaxy at larger *z* must have a higher (less negative) absolute magnitude *M* to be observed. Thus there are fewer galaxies per unit volume in the sample at larger *z*. The pair separations therefore tend to increase with *M* and *z* for both the red and blue samples. The effect on $L_0$ is small because we are effectively comparing the separation of red galaxies to blue and we chose equal numbers of red and blue galaxies (see Sec. 2 and 3), so that their average spatial densities were the same. It was not possible to evaluate any residual bias in the individual redshift slices because the range of absolute magnitudes was too small. Therefore this correction had to be determined using the complete sample.

The average *M* in the green band for each *z* slice was found to vary linearly with log *z* from −16.6 for 0.01<*z*<0.02 to −20.63 at 0.20<*z*<0.30. The galaxies in each *z* slice were divided into two halves with $M > \bar{M}$ and $M < \bar{M}$, where $\bar{M}$ is the average $M$. Then $L_0$ was calculated for each half and the slope $\Delta L_0 / \Delta M$ determined. The absolute magnitudes were determined using the apparent magnitude and $K$ corrections (Zehavi et al. 2002). An error-weighted fit to a constant $\Delta L_0 / \Delta M$ over the ten $z$ slices gave an average $\langle \Delta L_0 / \Delta M \rangle = 0.0405 \pm 0.0114$ with the error adjusted to give a $\chi^2/\text{dof} = 1$. Adjustments to the $L_0$ were then made using the value for the $0.08 < z < 0.10$ slice as an anchor, $\langle \Delta L_0 / \Delta M \rangle$ given above, and the difference in $\bar{M}$ for that $z$ slice and the $0.08 < z < 0.10$ slice. As a further check on the effect of decreasing spatial density with $z$, runs were made with only 25% of the galaxies that were chosen at random. At least 16 runs were made for each $z$ slice to reduce the statistical errors in $L_0$. The averaged $L_0$'s for these runs was consistently 3.6% higher (within statistical errors) than the normal runs for all $z$.

TABLE I. U-Z is the cut used to distinguish red and blue galaxies, $\bar{\xi}$ is the average correlation amplitude in bins 2–15, $L_0$ Corr is the correlation length after all corrections, StatErr is its statistical error, and TotErr its total error.

| Redshifts | U-Z | $\bar{\xi}$ | $10^3\ L_0$ Corr | StatErr | TotErr |
|---|---|---|---|---|---|
| 0.01-0.02 | 1.923 | 0.168 | 1.263 | 0.016 | 0.130 |
| 0.02-0.04 | 2.395 | 0.220 | 1.507 | 0.004 | 0.021 |
| 0.04-0.06 | 2.773 | 0.188 | 1.571 | 0.006 | 0.027 |
| 0.06-0.08 | 3.094 | 0.196 | 1.610 | 0.005 | 0.013 |
| 0.08-0.10 | 3.299 | 0.199 | 1.609 | 0.007 | 0.029 |
| 0.10-0.15 | 3.510 | 0.166 | 1.592 | 0.008 | 0.016 |
| 0.15-0.20 | 3.818 | 0.174 | 1.630 | 0.019 | 0.031 |
| 0.20-0.30 | 3.500 | 0.245 | 1.677 | 0.045 | 0.060 |
| 0.30-0.50 | 4.000 | 0.233 | 1.728 | 0.077 | 0.084 |
| 0.50-0.75 | 3.500 | 0.392 | 2.100 | 0.400 | 0.504 |

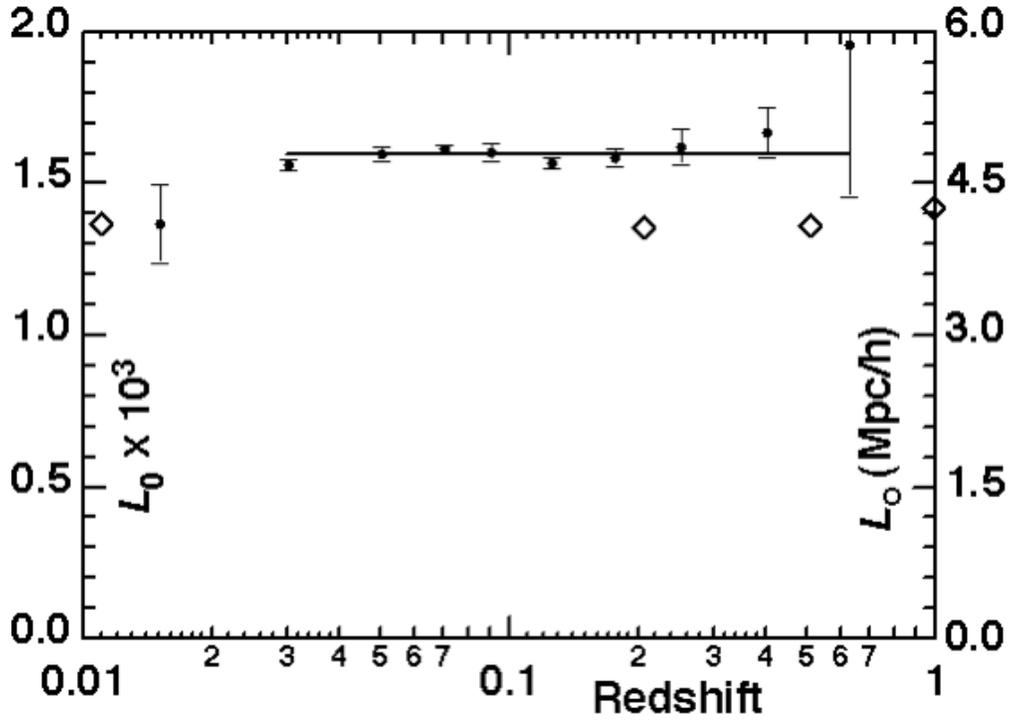

FIG. 3 $L_0$ vs. redshift after all corrections. The error bars are the total errors. The line is a fit to a constant, excluding the first point, that gives $\bar{L}_0 = (1.599\pm0.008)\times10^{-3}$ with a $\chi^2$/dof of 1.22. The scale on the right gives $L_0$ in units of Mpc/h. The diamonds show the values from the same analysis applied to the Millennium Simulation; the errors are smaller than the symbols.

*Variation of $L_0$ with cosmology.* – With $q_0 = 0.5$ rather than 0, $L_0$ changed about -2.5% at $z = 0.2$. This is not considered in the systematic errors since the same cosmological model should be used in any comparison with simulations.

## 5. RESULTS

The corrected $L_0$ are given in Table I along with error estimates and are plotted *vs.* redshift in Fig. 3. The horizontal line shows a least squares fit to a constant $L_0$ that yields $\bar{L}_0 = (1.599 \pm 0.008) \times 10^{-3}$ with a $\chi^2/\text{dof} = 1.22$. The first point is problematical because of the small volume involved, and it was not used in the fit. Using $H_0 = 100h$ km s$^{-1}$Mpc$^{-1}$ gives $\bar{L}_0 = 4.797 \pm 0.024$ Mpc/$h$.

Table I also gives $\bar{\xi}$, defined as the average correlation amplitude in bins 2–15. It is found to be remarkably consistent over the entire redshift range despite the huge variation of galaxy number density with redshift in the SDSS. This provides further evidence for the power of the red/blue difference technique in minimizing biases.

## 6. COMPARISON WITH THE MILLENNIUM SIMULATION

The Millennium Simulation (MS) is a very large simulation of the concordance ΛCDM cosmology. The Millennium Run follows the growth of dark matter structure from redshift $z = 127$ to the present using $10^{10}$ particles within a comoving box 500 $h^{-1}$Mpc on a side (Springel et al. 2005). In Croton et al. 2006 semi-analytic models are applied to the output of the Millennium Run to simulate the growth of galaxies and their central supermassive black holes and follow the detailed assembly history of each object. Public databases with the results are available at http://www.g-vo.org. The DeLucia2006a and DeLucia2006a_SDSS2MASS tables (de Lucia & Blaizot 2007) give positions, magnitudes, and other data for ~$10^7$ simulated galaxies with "snap-shots" in time ranging from $z$=127 to the present.[b] The analysis of the MS red-blue correlations was done with the same programs used for the SDSS with a subset of the MS galaxies chosen at random. The red-blue division was again made at the median in the $U$-$Z$ distribution. Variations of the $U$-$Z$ cut by 10% from the median caused changes <5% in the $L_0$. The $L_0$ were found to be very insensitive to the limiting magnitudes and the number of galaxies in the chosen subset. The resulting $L_0$ are plotted *vs*. epoch in Fig. 3 as the diamond points. The $L_0$ from the simulation show no significant variation with epoch/redshift, in good agreement with those from the SDSS. The MS ones are consistently about 10% lower than the SDSS. This suggests that the simulation parameters need to be tuned to improve the simulation, in particular, $\sigma_8$ the rms fluctuation in spheres of 8 $h^{-1}$ Mpc comoving radius.

In general the correlation *vs*. separation graphs from the MS are very similar to those for the SDSS, which are illustrated in Fig. 2. For epochs <0.4 the correlation amplitudes $\bar{\xi}$ were ~0.28, somewhat larger than those in Table I . These decreased to ~0.20 at epoch 1.0. A detailed comparison of spatial fluctuations in SDSS/MS correlation lengths and amplitudes will be forthcoming.

## 6. DISCUSSION

The correlation lengths found here are roughly consistent with other data on cluster correlation lengths (Zehavi et al. 2002; Davis & Peebles 1983; Carlberg et al. 2000), though the sizes are definition dependent. There is no evidence for evolution of cluster size with redshift out to $z$ =0.5, in agreement with the Millennium Simulation. In contrast, Zehavi et al. (2005), using the standard $\xi(r) = (r/r_0)^{-\gamma}$ fit to SDSS data, find that $r_0$ increases from 2.83±0.19 Mpc/$h$ for R-band luminosity $M_r \approx -17.5$ and $z \approx 0.02$ to 10.00±0.08 Mpc/$h$ for $M_r \approx -22.5$ and $z \approx 0.17$. Carlberg et al., analyzing the CNOC2 high-luminosity sample with a luminosity compensated absolute magnitude, find that $r_0$ <u>decreases</u> slightly from 4.75±0.05 Mpc/$h$ for $z \approx 0.10$ to 4.26±0.18 Mpc/$h$ for $z \approx 0.49$.

The amplitude of the correlation ($\bar{\xi}$ in Table I) is a measure of the clustering strength and is remarkably constant over the whole $z$ range. This shows that the cluster number density remains substantially constant out to $z$~0.5, consistent with the MS. The shape of $\xi(r)$ also provides information on the density profile of the clusters.

Aside from possible variations of cluster size with epoch, this technique provides a sensitive probe of possible variations in cluster size or number density with angle or position. Despite the piecemeal SDSS coverage in $\delta$ of the hemisphere toward $RA=0°$, the $L_0$ were found to be the same as the $RA\sim180°$ hemisphere within statistical errors that were about 5 times larger. A detailed analysis of the distribution of average cluster size and number density will be required to determine if these are consistent with the fluctuations to be expected in the standard $\Lambda$CDM model. The correlation dependence on separation, illustrated in Fig. 2, serves as a proxy for the mass density profiles and this provides another way to test $\Lambda$CDM. In addition, the improved use of the existing galaxy survey data provides a better way to search for violations of spatial homogeneity (Alnes, Amarzguioui, and Grøn 2006), such as the existence of a preferred axis (Schwarz et al. 2004) or of a large void (Clifton, Ferreira and Land 2008).

---

[b] I shall refer to these snapshots as epochs to distinguish them from SDSS redshifts that correspond to distances as well as lookback times.

The technique of using the red *vs.* blue galaxy comparison to determine average cluster sizes has the advantages of great statistical power as well as being almost free of selection biases. I find that the average cluster size and their number density are nearly constant out to redshifts of 0.5, in very good agreement with results using the same technique applied to the Millennium Simulation. With deeper surveys and more careful redshift measurements it will be possible to extend the technique to considerably larger *z*. This will make it competitive with Type Ia supernovae in establishing an accurate cosmic distance scale. (See, for example, Frieman, Turner & Huterer 2008 for a recent discussion.) The definitions of cluster size and other parameters used here can be readily incorporated into simulations so that more detailed comparisons of the data with cosmological models can be made.

This analysis would not have been possible without the dedicated efforts of the SDSS collaboration. The Millennium Run simulation was carried out by the Virgo Supercomputing Consortium at the Computing Centre of the Max-Planck Society in Garching. The semi-analytic galaxy catalog is available at http://www.mpa garching.mpg.de/galform/agnpaper. I am grateful to G. Lemson for help in using the MS and A. Evrard for helpful discussions of galaxy clustering.